\documentclass[12pt]{article} 
 
\usepackage{amssymb,cite} 
\usepackage{graphicx} 
\usepackage{latexsym} 
\usepackage{epsfig,epic,amsfonts} 
\usepackage{wrapfig} 
\usepackage{bm}
\usepackage[ansinew]{inputenc}
\usepackage{rotating}
\usepackage{euscript}
 
\def\thefootnote{\fnsymbol{footnote}}

\newcommand{\eq}{\begin{equation}} 
\newcommand{\en}{\end{equation}} 
\newcommand{\be}{\begin{equation}} 
\newcommand{\ee}{\end{equation}} 
\newcommand{\eqa}{\begin{eqnarray}} 
\newcommand{\ena}{\end{eqnarray}} 
\newcommand{\ba}{\begin{eqnarray}} 
\newcommand{\ea}{\end{eqnarray}}

\newcommand{\ZZ}{\hbox{{\rm Z{\hbox to 3pt{\hss\rm Z}}}}}

\newcommand{\EQ}{\begin{equation}} 
\newcommand{\EN}{\end{equation}} 
\newcommand{\bea}{\begin{eqnarray}} 
\newcommand{\eea}{\end{eqnarray}}

\begin{document} 
\begin{titlepage} 
\vskip0.5cm 
\begin{flushright} 
DFTT 52/09\\ 
\end{flushright} 
\vskip0.5cm 
\begin{center} 
{\Large\bf  Nucleation dynamics in 2d cylindrical Ising models and chemotaxis } 
\end{center} 
\vskip1.3cm 
 
\centerline{C.Bosia$^1$$^,$$^2$, M.Caselle$^1$$^,$$^2$ and D.Cor\'a$^1$$^,$$^2$$^,$$^3$}    
    
 \vskip0.4cm    
 \centerline{\sl  $^1$ Dipartimento di Fisica    
 Teorica dell'Universit\`a di Torino and I.N.F.N.,}    
 \centerline{\sl via P.Giuria 1, I-10125 Torino, Italy}   
 \centerline{\sl  $^2$ Center for Complex Systems in Molecular Biology and Medicine,}    
 \centerline{\sl University of Torino, Via Accademia Albertina 13, I-10100 Torino, Italy}  
 \centerline{\sl  $^3$ Current address: Systems Biology Lab, Institute for Cancer Research and Treatment (IRCC),}    
 \centerline{\sl  School of Medicine, University of Torino, Str. Prov. 142 Km. 3.95, I-10060 Candiolo, Torino, Italy}
 \centerline{\sl    
e--mail: \hskip 1cm (cbosia)(caselle)(cora)@to.infn.it}    
 \vskip1cm    
    
\begin{abstract}
The aim of our work is to study the effect of geometry variation on nucleation times
and to address its role in the context of eukaryotic chemotaxis 
(i.e. the process which allows cells to identify and follow a gradient 
of chemical attractant).
As a first step in this direction 
we study the nucleation dynamics of the 2d Ising model defined on a cylindrical
lattice whose radius changes as a function of time. 
Geometry variation is obtained by changing the relative value of the couplings
between spins in the compactified (vertical) direction 
with respect to the horizontal one. This allows us to keep the lattice size
unchanged and study in a single simulation the values of the compactification
radius which change in time. We show, both with theoretical arguments and numerical
simulations, that
squeezing the geometry allows the system to speed up nucleation times even in
presence of
a very small energy gap between the stable and the metastable states.
We then address the implications of our analysis for directional chemotaxis. 
The initial steps of chemotaxis can be modelled as a nucleation process occurring on
the cell membrane
as a consequence of the external chemical gradient (which plays the role of energy
gap between the stable and metastable
phases). In nature most of the cells modify their geometry by extending
quasi-onedimensional protrusions (filopodia) so as to enhance their
sensitivity to chemoattractant. Our results show that
this geometry variation has indeed the effect of greatly decreasing the timescale of
the nucleation process even in presence of very small amounts of
chemoattractants.
\end{abstract}    
\end{titlepage}    

\setcounter{footnote}{0} 
\def\thefootnote{\arabic{footnote}} 
\section{Introduction} 
\label{introsect} 

The decay of the metastable state at a first order phase transition is a rather well understood problem both 
in statistical mechanics and in quantum field theory. It attracted much interest in these last years due to its role in various different
physical contexts ranging from the condensed matter applications to the study of the 
deconfinement transition in QCD or of the inflationary mechanism in cosmology. 
The crucial problem in all these contexts is the evaluation of the lifetime of the metastable state from the knowledge of the microscopic interactions of the system
and of the details of the non-equilibrium dynamics which drives it toward the true equilibrium state. If the Hamiltonian of the system only contains
 short range interactions then, as a general result, metastable states will always decay but their lifetime can vary by orders of magnitude.
A simple but powerful 
effective model for the description of metastable decay is the so called droplet model which traces back to the seminal papers by Becker and
D\"{o}ring and Zeldovich ~\cite{BEDO35,ZELD43}, and was later reinterpreted and extended using a field-theoretic 
language by Langer and Coleman and Callan ~\cite{Langer,COLE77,CALL77} (for a review see for instance ~\cite{rg94}). The main result of this approach is that the lifetime of
the metastable state depends exponentially on the free energy $F_c$ 
of the so called ``critical droplet'' whose size $R_c$ can be evaluated by comparing the bulk and perimeter
contributions to the free energy of the droplet. 
The weak metastability regime (in which we are interested in this paper) is reached when the $F_c$ is much larger than
the temperature of the system (or than the Plank's constant in field theoretic applications).

In this ``classical'' framework the system is assumed to be of infinite
extension and the background spacetime geometry is assumed to be flat and fixed. The picture becomes much more interesting if these two constraints are relaxed. 
When the system has a finite size $L$, comparable with the critical radius $R_c$, then the competition between these two scales may induce very different behaviours
and creates a combination of different dynamical regimes~\cite{rtms}. In particular for $L<<R_c$, the limiting scale of the nucleation process becomes 
the ratio $L/\xi$ where $\xi$ is the bulk correlation length of the system. 
When the background geometry is allowed to fluctuate
then the exponential regime itself is abrogated and one finds instead a power law dependence of the decay rate ~\cite{Zamolodchikov:2006xs} on the energy gap. 

The aim of our paper is to show, following the above observations,  that by 
acting on the background geometry one can indeed modify the decay rate of the system and reach, 
even for very small energy gaps between the stable and metastable phases, very short values for the
lifetime of the metastable phase.  

Besides its intrinsic interest, 
this feature may be of some importance for biological applications. Nucleation theory, among several other biological applications, plays
an important role in eukaryotic chemotaxis (the process which allows eukaryotic cells to identify and follow a gradient of chemical attractant).
Chemotaxis is mediated by two key enzymes: $PI3K$ and $PTEN$. The standard way to
model the process is to assume that the cell membrane, as a consequence of the external chemical gradient, undergoes a phase ordering process
reaching a stable phase separation regime  between a $PI3K$-rich phase (toward the chemoattractant)
and a $PTEN$-rich phase (opposite to the chemoattractant)~\cite{g3,g2,g1}. In this picture the chemical gradient plays the role of energy gap between the stable and metastable
phases. The limiting scale of this process is the cell size. Gradients smaller than a critical value would require critical droplets larger than the cell size to
drive the process and this would lead, depending on the details of the model, to a random orientation of the two phases or to 
a dramatic increase in the timescale of the process. The way cells adopt to overcome this problem and enhance their
sensitivity to chemoattractant is to modify their geometry by extending quasi-onedimensional protrusions (filopodia). This change of background geometry
is exactly the type of process that we plan to study in this paper.

All along the paper we shall use the two dimensional Ising model as 
prototypical example of a statistical system with short  range interactions showing metastable phases when an external magnetic field is applied in the broken
symmetry phase of the model. Besides this paradigmatic role there are two other important reasons which led us to concentrate in the Ising model.

The first one is that it has been recently shown in  \cite{fcgc07} that the chemotacting model described above can be mapped into a suitable
generalization of the 2d Ising model. The standard Ising model is obtained in the limit of infinite amount of cytosolic $PTEN$ and $PI3K$ enzymes 
(we shall discuss this mapping in some details below).

The second reason is somehow more technical but of similar importance for our purposes. The 2d Ising model
can be exactly solved (and the correlation length can be exactly  evaluated) even for asymmetric values of the coupling constants. This fact will
play a crucial role in the protocol that we shall use to modify the background geometry of the model. In fact it will allow us to modify the geometry by simply
acting on the coupling constants of the model without changing the structure of the
2d lattice (which would require a very sophisticated and inefficient software).

This paper is organized as follows. In the next section we shall briefly discuss the chemotacting model mentioned above.
In sect.3 we shall introduce the Ising model with asymmetric couplings and briefly discuss its phase diagram.
Sect.4 will be devoted to a brief review of the classical nucleation theory (CNT) with a particular attention to the interplay between the two main 
scales of the problem:
the system size $L$ and the critical droplet radius $R_c$. In sect.5
we shall discuss the procedure we adopted to vary the geometry of the system and the results of
our simulations while the last section will be devoted to a few concluding remarks.

\section{A simple chemotacting model and its Ising-like realization} 
\label{sect4}

The link between  nucleation dynamics in  2d Ising models and chemotaxis is based 
on the observation that chemotaxis requires as a preliminary step a phase separation process on the membrane of the cell~\cite{g3,g2,g1}.

When a chemoattractant is switched-on in the external environment of the cell, owing to the interplay of the two enzymes $PTEN$ (phosphatase and 
tensin-homolog) and $PI3K$ (phosphatidylinositol 3-kinase), the phospholipids $PIP_2$ (phosphatidylinositol bisphosphate) and $PIP_3$ 
(phosphatidylinositol trisphosphate) are interconverted: $PI3K$ catalyzes $PIP_2$ phosphorylation and $PTEN$ catalyzes $PIP_3$ dephosphorylation.
While the phospholipids are bound to the inner surface of the membrane, the enzymes can freely 
diffuse in the cytoplasm, becoming active if absorbed on the membrane. When it happens, it is observed the formation of two patches 
in complementary domains, rich in $PIP_2$ and $PIP_3$ respectively. In particular, the $PIP_3$-rich patch is localized on the side of the membrane 
exposed to the highest concentration of chemoattractant (the leading edge) while the $PIP_2$ rich-one on the other side (the rear edge). Similarly, a 
phase separation also involves cytoplasmatic localization of the enzymes $PI3K$ and $PTEN$. 

Such spatial-organization phenomenon may be seen as a self-organized phase ordering process, where the cell state - driven by an external field - decays
into two spatially localized chemical phases, thus defining a front and a rear. This process lies at the heart of directional sensing, finally leading to
cell movement toward the maximum of the attractant gradient. 

Following ~\cite{fcgc07} the dynamics of the phase separation process occurring in the cell membrane
can be modeled with a suitable generalization of the 2d Ising model. 
The cell membrane is represented by a square lattice of side $L$ with periodic boundary conditions and the spin variables $S_i = +1$ 
($-1$) are the enzymes $PI3K$ ($PTEN$). The Hamiltonian of the model is composed by three terms: 
a short-range (i.e. involving nearest-neighborhood sites) ferromagnetic coupling describing the attractive interaction between enzymes, an external site-dependent magnetic 
field mapping the effect of the attractant and a long-range antiferromagnetic interaction term which keeps into account
the finiteness of the cytosolic enzymatic reservoir.

From the ratio of probabilities that $PI3K$ and $PTEN$ enzymes bind to the site $i$, it is possible to trace the energy 
difference between states $S_i = +1$ and $S_i = -1$,
\eq
\label{eq:delta}
\Delta H = -2J \sum_{j \in \partial i} S_j - 2 h_i + 2 \lambda m
\en
(where $j \in \partial i$ are the j nearest-neighborhood sites of the site i) which can be reinterpreted as the variation of the Hamiltonian
\eq
\label{eq:Hnuova}
H = - J \sum_{<i,j>} S_i S_j - \sum_{i} h_i S_i + \frac{\lambda}{N}\sum_{i<j} S_i S_j \, ,
\en  
where J is the ferromagnetic coupling constant, the first sum is over all the nearest-neighborhood sites ($<i,j>$), $h_i$ is the external 
site-dependent field representing the chemoattractant, $N$ is the total number of enzymes $PTEN$ and $PI3K$
and $\lambda$ is a constant weighing the antiferromagnetic term with respect to the ferromagnetic one.
The standard Ising model is recovered in the $N \rightarrow \infty$ limit, i.e. for an infinite enzymatic reservoir. However it is important to stress that
this is a non trivial limit.
In fact due to the additional antiferromagnetic term the generalized 2d Ising model behaves in a quite different way with respect to the standard Ising model.
In the broken symmetry phase it shows self-tuned phase separations also for very small amounts of gradient. However if the gradient is too small the two phases are
randomly distributed on the lattice (i.e. the cell membrane) and chemotaxis cannot occur. In this generalized Ising model
the absence of correlation between the spatial orientation of the two phases and the external field 
plays the role of the exponential increase of the mestable state lifetime in the standard Ising model. In this paper we decided to concentrate in the 
$N \rightarrow \infty$ Ising limit since it allows a rigorous theoretical description. We plan to devote a forthcoming paper to a detailed numerical analysis of its
finite $N$ generalization.

In the above model the geometry plays no role: the lattice is a flat two dimensional 
torus which is assumed to be a reasonable approximation of a spherical cell membrane. However this
is a very crude approximation from a biological point of view.
Indeed in the chemotactic process, the geometry of the cell plays a pivotal role: 
signal sensing takes place as the leading edges of nearly one-dimensional finger-like 
projections (called {\itshape filopodia}) protrude from the cellular body. Such structures, as sophisticated antennas, allow the cell to sense even 
subtle concentrations of external signal, finally enabling motion (see ~\cite{Bray, Guillou, Borm}). Since filopodia are dynamic structures appearing to sense the signal when the source 
of chemoattractant
is far from the cell, the membrane is continually forced to change its geometry, squeezing and splitting its leading edge. 
Moreover, since in a nearly one-dimensional geometry the interface tension between two different phases is lower than in a spherical geometry, even
a very small gradient of external field can drive the coarsening toward its maximum. If the geometry fluctuates, the thin fingers developed from the 
leading edge can sense small amounts of gradient thanks to the rapid formation of clusters rich in $PIP_3$ in the free extremities (and $PIP_2$ in their
complementary share) and then maintain the phase separation in the spherical geometry once they are reabsorbed by the membrane. 

As mentioned in the introduction the main goal of our work is to try to include these effects in the model by simulating the formation of a quasi 
onedimensional filopodium and then its reabsorption within the cell membrane.

\section{Ising model with asymmetric couplings} 
\label{sect2} 
\subsection{General setting} 
The 2d Ising model with asymmetric couplings is defined by the following Hamiltonian ~\cite{baxter}.
\eq
\label{eq:H-asymm-coupl}
H = -J_x \sum_{<i,j>} S_i S_j - J_y \sum_{<i,k>} S_i S_k \, ,
\en 
where $J_x$ and $J_y$ denote the horizontal and vertical couplings respectively, the sums are over the horizontal ($<i,j>$) and vertical ($<i,k>$) 
nearest-neighborhood sites and the spins $S_i$ take as usual the values  $S_i = \pm 1$. 
We shall denote in the following with $L_x$ and $L_y$ the sizes of the lattice in the 
horizontal and vertical directions respectively.

The corresponding partition function is
\eq
\label{eq:Z-H-asymm-coupl}
Z_N = \sum_{S} \exp \left[ K \sum_{<i,j>} S_i S_j + W \sum_{<i,k>} S_i S_k \right]
\en
with $K = \frac{J_x}{k_B T}$ and $W = \frac{J_y}{k_B T}$ , $k_B$ is the Boltzmann's constant and $N = L_x L_y$ is the number of sites of the lattice. 
\\
The Kramer-Wannier duality holds also for this model and has the effect of relating the two couplings
(for a review see for instance~\cite{baxter, fisher-burford}).
The duality relations are:
\eq
\mbox{sinh} \left( 2 K^\ast \right) \mbox{sinh} \left( 2 W \right) = 1 \; , \; \mbox{sinh} \left(2 W^\ast \right) \mbox{sinh} \left( 2 K \right) = 1 \; ,
\label{u1}
\en
or equivalently: $\mbox{tanh} \, K^\ast = e^{-2W}$, $\mbox{tanh} \, W^\ast = e^{-2K}$.
Denoting with  $\psi \left( K,W \right)$  the free energy of the model we have:
\eq
\label{eq:free-energy}
\psi \left( K^\ast , W^\ast \right) = \psi \left( K,W \right) + \frac{1}{2} \ln{ \left( \mbox{sinh} \left( 2K \right) \mbox{sinh} \left( 2W \right) \right)} \; ,
\en
from eq.(\ref{u1}) we can easily obtain the selfdual line which in this case is also the critical line of the model:
\eq
\label{eq:critical-line}
\mbox{sinh} \left( 2K \right) \mbox{sinh} \left( 2W \right) = 1 \; .
\en 
For $K = W$ this equation becomes the well known selfdual condition which allows to obtain the critical temperature of  the standard 2d Ising model.

All the points along this critical line correspond to the same universality class and are described in the continuum limit by the same 2d conformal field theory
(CFT) with $c=1/2$ (where $c$ is conformal anomaly of the underlying CFT).

Outside the critical line the equation
$ \mbox{sinh} \left( 2K \right) \mbox{sinh} \left( 2W \right) = k$, with $k\not= 1$ constant, defines an infinite set of curves of constant temperature $T$ in the $K,W$ plane. In particular
for  $k < 1$ we have the  $T > T_c$ (high-temperature) phase 
while for $k > 1$ we have  $T < T_c$ (low-temperature) phase, where $T_c$ is the critical temperature of the symmetric Ising model.
In the following we shall be interested in the low T phase (i.e. $k>1$) in which the $Z_2$ symmetry of the model is spontaneously broken.

For reasons which will be clear below we choose different boundary conditions in the two directions: periodic b.c. in the vertical ($y$-axis) 
direction and free b.c. in the horizontal ($x$-axis) direction. The lattice thus acquires a cylindrical geometry\footnote{In the quantum field theory (QFT) 
interpretation of the
model this choice of boundary conditions corresponds to the so called ``finite temperature regularization'' of the QFT. In this framework the model describes in the
continuum limit a one dimensional QFT in contact with a heat bath whose temperature (which has nothing to do with the temperature which appears in the Hamiltonian of
the model) is related to the
inverse of the lattice size in the compactified
direction (for a review see for instance \cite{Billo:1996pu}).}. 

The most interesting feature of this model is that it can be solved exactly for any value of $K$ and $W$ in absence of an
external magnetic field (see for instance \cite{baxter}). Moreover an exact expression for the correlation length in the two spatial directions can be obtained in
the high temperature phase, by exact resummation of the strong coupling expansion~\cite{fisher-burford}: 
\eq
\label{eq:csi}
\xi_x = - \frac{1}{\ln{ \left[ \mbox{tanh}\, K \left( \frac{1+ \mbox{tanh}\, W}{1- \mbox{tanh}\, W} \right) \right]}}
\;
,
\;
\xi_y = - \frac{1}{\ln{ \left[ \mbox{tanh}\, W \left( \frac{1+ \mbox{tanh}\, K}{1- \mbox{tanh}\, K} \right) \right] }} \; .
\en
\\
From these expressions, using the duality transformation one can obtain the analogous values for the correlation length in the low T phase. This derivation however
is not completely trivial. One should keep into account that in the vicinity of the critical point low and high T correlation lengths are related by an universal
amplitude ratio $R_\xi\equiv \xi_+/\xi_-$ where $\xi_+$ and $\xi_-$ denote the correlation lengths above and below the critical point.
In the case of the 2d Ising universality class the value of $R_\xi$ is exactly known to be $R_\xi=2$. This value keeps track of the fact that the spectrum of the 
underlying quantum field theory is very different in the two phases.

These correlation lengths will play an important role in the following since they will allow us to evaluate the effective geometry of the model 
in the continuum limit. 

\subsection{Asymmetric couplings and lattice deformations } 

The most interesting feature of the Ising model with asymmetric couplings is that by changing the couplings along a line of constant
absolute temperature, i.e. at fixed $k$, we can effectively modify the aspect ratio of the lattice. This is indeed a standard tool in finite temperature quantum field
theory ~\cite{Billo:1996pu}. In fact it is well known in this context that we may obtain two physically equivalent regularizations of the model 
by squeezing the radius of the cylinder or, in a completely equivalent way, keeping the radius fixed and suitably increasing the coupling in
the compactified (vertical) direction and simultaneously decreasing the coupling in the orthogonal (horizontal) directions. The exact amount of this change of the
couplings can be determined using as physical scale the correlation length of the model. As a consequence of the changes in the couplings the vertical (i.e. along
the compactified direction) correlation length $\xi_y$ increases while the horizontal one $\xi_x$ decreases. 
Since the only other scale of the problem is the circumference of the
cyilinder $L$, models with the same ratio $L/\xi_y$ are physically equivalent. We can thus trade a change in the lattice size in the compactified direction with a
change in the value of the vertical coupling. This equivalence is routinely used in lattice gauge theories and 
has been recently confirmed also in the context of the 2d Ising model with a set of high  precision simulations of the Binder cumulant \cite{s09}.
This equivalence will allow us to construct simulation protocols in which we can change in a continuous way the radius of the cylinder (keeping the lattice size
fixed and changing the couplings) thus mimicking the elongation of eukaryotic cells along the gradient of a chemoattracting potential and even the extreme situation
of the formation of quasi one-dimensional filopodia. 

\subsection{External magnetic field and metastability } 

\begin{sidewaysfigure}
  \centering
  \includegraphics[scale=1.3]{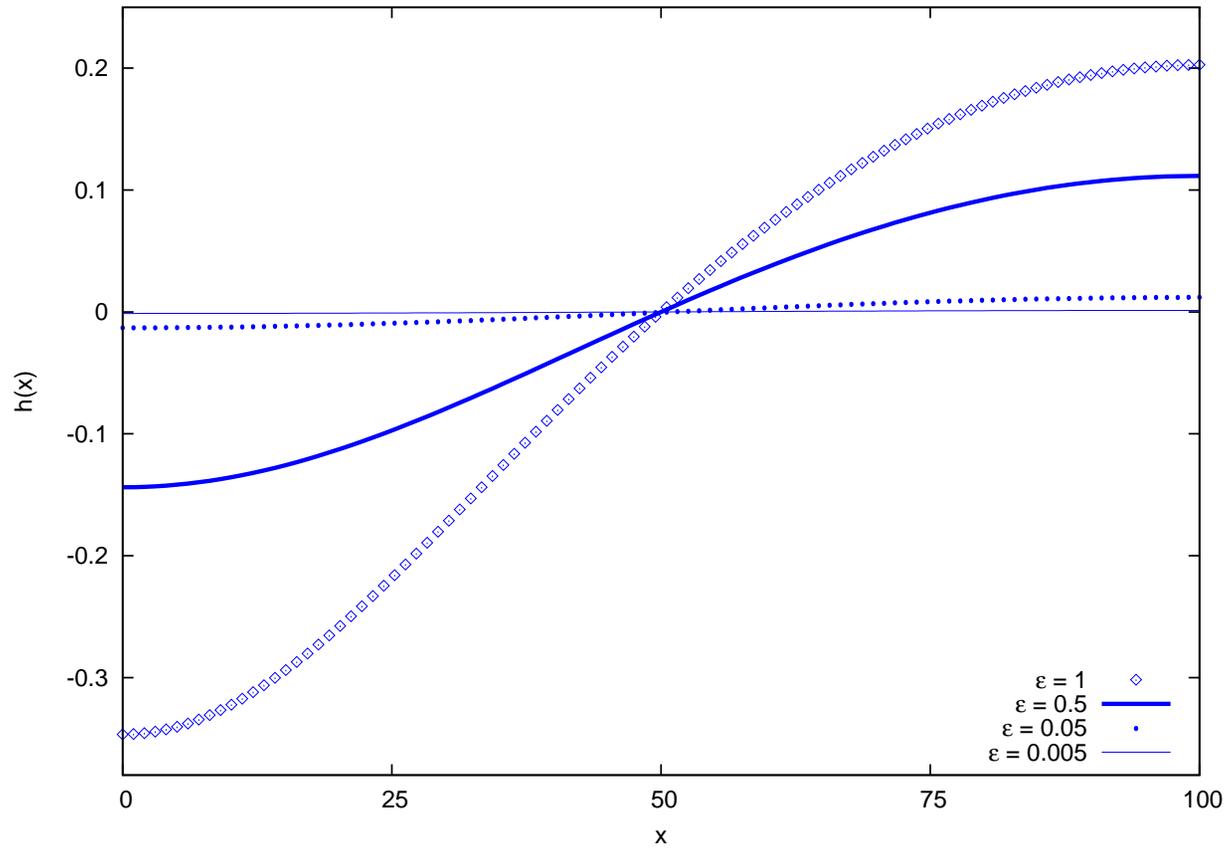}
  \caption{(Color online). Site-dependent external magnetic field for several values of $\epsilon$ with $c=1$. On the vertical axis is reported the site-dependent
external magnetic field $h(x)$ while on the the horizontal axis $x$ runs on the $L_x$ lattice side.\label{campo}}
\end{sidewaysfigure}

In order to study the metastable behaviour of the model we must add to the Hamiltonian an external magnetic field. In order to mimic an external chemotactic 
gradient,
instead of choosing as usual an uniform external field, we choose to couple the system to a site dependent field whose shape is plotted in fig. (\ref{campo}).

The Hamiltonian considered is therefore the following:
\eq
\label{eq:H-asymm-coupl+field}
H = - J_x \sum_{<i,j>} S_i S_j - J_y \sum_{<i,k>} S_i S_k - \sum_{i} h_i S_i
\en
which leads to the partition function
\eq
\label{eq:Z-H-asymm-coupl+field}
Z_N = \sum_{S} \exp \left[ K \sum_{<i,j>} S_i S_j + W \sum_{<i,k>} S_i S_k + \beta \sum_{i} h_i S_i \right] .
\en
\\
$N$, $K$ and $W$ as well as $<i,j>$ and $<i,k>$ are defined as above, $\beta = \frac{1}{k_B T}$ and 
\eq
h_i = \frac{1}{2} \ln \left( 1 + c_i \right) - \frac{1}{2} \ln \left( 1 + c \right) 
\label{eq:h1}
\en
is the external magnetic field, where 
\eq
c_i = c \left( 1 - \epsilon \mbox{cos} \left( \frac{\pi x_i}{L_x} \right) \right) .
\label{eq:h2}
\en
$c$ is a constant, $\epsilon$ defines the gradient intensity of the magnetic field and $L_x$ is the side of the lattice along the x-axis.
With this definition the stable state is characterized by two phases (with minus sign for low values of $x$ and plus sign for large values of $x$) separated by an
interface while the metastable state, whose decay properties we shall study in the following, will be defined as the state in which all the spins are pointed in the
$-1$ direction.

\section{Classical Nucleation Theory} 
\label{sect3} 

In this section we
shall first briefly summarize the classical nucleation theory in the case of the Ising model defined 
on a square lattice of infinite size, then in the second part of the section we shall see the effect of introducing a finite size scale $L$ comparable with the
critical radius $R_c$. Most of the results reported in this section can be found in standard textbooks. For a detailed discussion see \cite{rtms}.

\subsection{Infinite systems}
\label{sect3a}

In the Classical Nucleation Theory the decay of the metastable state is controlled by three main physical scales: the correlation length $\xi$
which sets the typical 
mean size of the droplet which are created in the system by thermal fluctuations, 
the critical droplet radius $R_{c}$  and the size $R_0$
to which one droplet can grow before it is likely to meet another one. CNT usually assumes $\xi<<R_c<<R_0$ and from simple thermodynamic arguments allows to obtain
reliable estimates for the nucleation rate and for the mean lifetime of the metastable state. Let us briefly remind the main steps of this calculation.

Let us first evaluate $R_c$. To keep the analysis as simple as possible we shall assume the droplet to be of circular shape with area $\pi R^2$ and perimeter 
$2\pi R$. It is easy to show that the analysis holds for any generic shape.
The free energy of the droplet is
\eq
\label{eq:1}
F(R) = 2 \pi  R {\sigma_0} - \pi R^2 \Delta E \;,
\en
where $\Delta E$ is the difference in bulk free-energy density between the
metastable and stable states and ${\sigma_0}$ is the surface tension. 
In the Ising case in which we are interested the energy gap is proportional to the absolute value of the magnetic field $|H|$. For not too large magnetic fields 
a good approximation for $\Delta E$ turns out to be
$\Delta E=2MH/\beta$ where $M$ is the spontaneous magnetization.

The probability of a fluctuation of this size is given by

\eq
\label{eq:2}
P\sim e^{-\frac{F(R)}{k_B T}} \; .
\en

Applying standard droplet theory arguments the critical radius is the value which minimizes (\ref{eq:2})

\eq
\label{eq:3}
R_{\rm c}(T,H) = \frac{{\sigma_0}}{\Delta E} .
\end{equation}
The nucleation rate (i.e. the probability of a critical fluctuation)
 is dominated by the free energy of the critical droplet $F_c=\frac{\pi\sigma^2_0}{\Delta E}$ :
\eq
\label{eq:4}
I(T,H)\sim e^{-\frac{\pi\beta\sigma^2_0}{\Delta E}}\sim e^{-\frac{\pi\beta^2\sigma^2_0}{2M|H|}}\equiv e^{-\frac{\Gamma}{|H|}}
\en

where we have introduced the constant $\Gamma=\frac{\pi\beta^2\sigma^2_0}{2M}$ which keeps track of all the microscopic details of the model.

Let us discuss a few important features of this classical result.
\begin{itemize}
\item
Eq.(\ref{eq:4}) is only the first order in the semiclassical approximation ~\cite{Langer,COLE77,CALL77}. The next to leading order can be evaluated using quantum field theory methods and leads
to a prefactor whose power can be predicted analitically and agrees very well with high precision Montecarlo simulations (see for instance ~\cite{Munster}).
We shall not further discuss this issue in
the present paper since for our purposes the  semiclassical result will be enough.

\item

In eq.(\ref{eq:4}) the information about the first scale we have in the game, the correlation length of the model $\xi$, is hidden in the constant $\Gamma$. More
precisely in the scaling region the string tension is related to $\xi$ by the universal amplitude ratio $R_\sigma\equiv \sigma_0\xi$ \cite{pv}. 
The requirement $\xi<< R_c$ is thus translated into the
condition $\Gamma>>|H|$ and allows us to have a precise estimate of the values of $H$ for which the CNT can be used. At the same time the $\Gamma>>|H|$ condition
implies that we should expect exponentially long nucleation times.

\item

In the above derivation we assumed a single droplet approximation. This is a reliable choice as far as critical 
droplets are far enough from each other. This is the
physical meaning of the condition $R_c<<R_0$ mentioned above. It can be shown (see for instance ~\cite{rtms}) that $R_0$ increases exponentially with $1/|H|$ thus,
since $R_c \sim 1/|H|$, there will always be a value of $H$ below which this approximation is correct.

\end{itemize}

\subsection{Finite size systems}

If we consider a system of finite size $L$ there will always be a value of $|H|$ small enough such that $R_c>L$. In this limit the system behaves as along the
coexistence line i.e. as if $|H|$ would be negligible with respect to the other scales. Thus we can neglect the area term in eq.(\ref{eq:1}) and the droplet free
energy becomes proportional to $R$. The saddle point analysis discussed above does not hold any more and one can show that instead the metastable lifetime
increases exponentially with the system size $L$. To address this point let us define as $\phi$ the fraction of the system occupied by the droplet, i.e. 
(assuming again the simplified 
picture of a spherical droplet) $\phi=\pi R^2/L^2$. Then we can rewrite the free energy as 
\eq
\label{eq:6}
F(R) \sim 2 \sqrt{\pi \phi}  L {\sigma_0} \; .
\en

As above we can trade the line tension $\sigma_0$ for the correlation length using the universal amplitude ratio $R_\sigma$ and rewrite
$F$ as
\eq
\label{eq:7}
F(R) \sim A  \frac{L} {\xi}
\en
with $A=2 \sqrt{\pi \phi} R_\sigma $ which, 
if we are interested in the free energy of a droplet which covers a finite fraction of the system,
is a constant of order unity. Thus we see that the parameter which governs the droplet formation is actually the ratio $L/\xi$ and that the 
lifetime of the metastable state increases exponentially with $L/\xi$.

In particular, if we are interested in a cylindrical geometry, the limiting scale will be the radius of the cilinder and, in case of asymmetric couplings, the
correlation length in eq.(\ref{eq:7}) will be the correlation length along the compactified direction.

\section{Simulation setting and results} 
\label{sect5} 
Simulations were performed using a standard Glauber dynamics in order to mimic the local interactions of enzymes on the membrane surface \cite{fcgc07}.

We studied the model on a square lattice of sizes  $L_x = L_y = L =100$. The system was initialized setting all the spins to the  $S_i = -1$ value. 
The boundary conditions were chosen to be  periodic in the $y$-axis and free in the $x$-axis.
For definiteness we set in eq.(\ref{eq:h1},\ref{eq:h2}) $c = 1$ and  considered six values of gradient intensity, ranging from $\epsilon = 5$ to 
$\epsilon = 0.0005$ (see tab.\ref{tab2}). 

\begin{sidewaysfigure}
  \centering
  \includegraphics[scale=1.3]{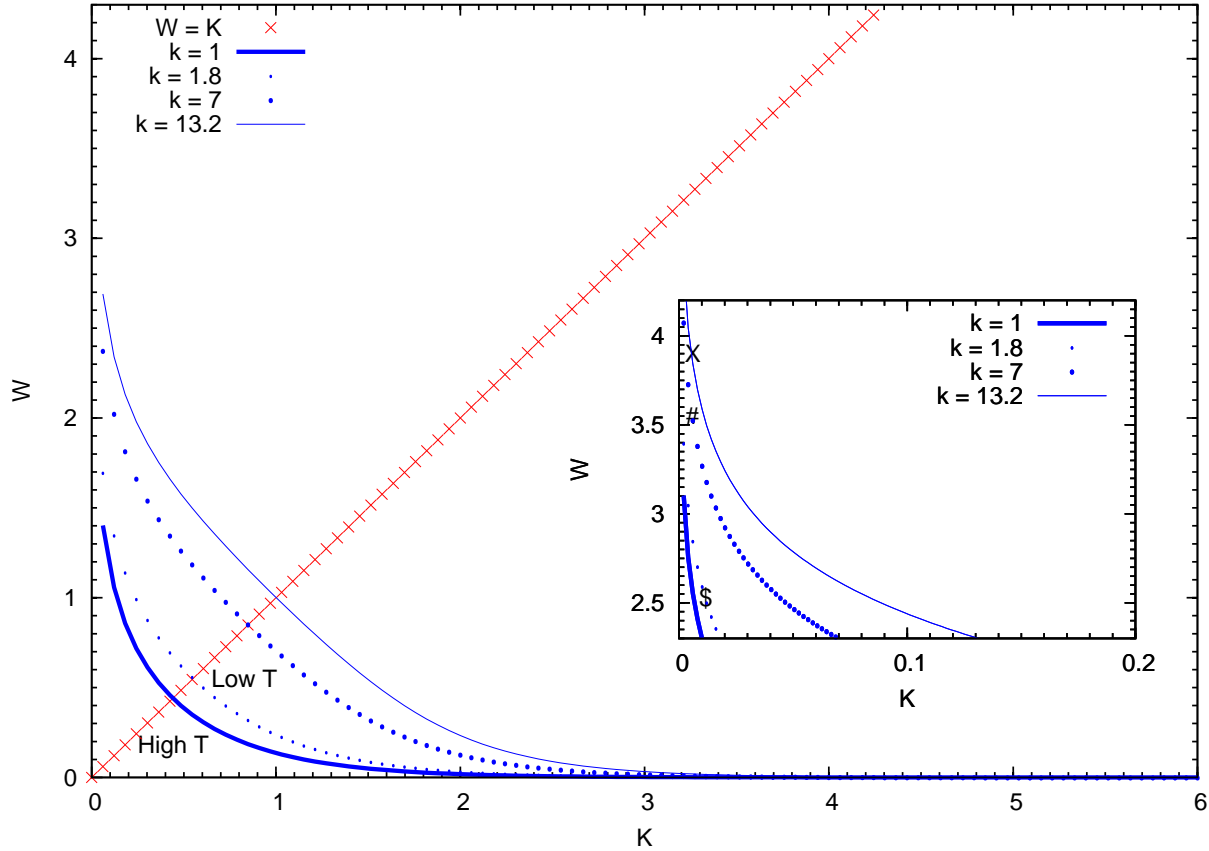}
  \caption{(Color online). $W=\frac{1}{2} \mbox{asinh} \left( \frac{k}{\mbox{sinh} \left( 2K \right)} \right)$ for several values of $k$. On the $y$ and $x$-axis
are reported $W$ and $K$ respectively. In the inset the points (K;W) 
corresponding to $\xi_y = \frac{L_y}{2}$ are plotted: the \$ symbol refers to the point $(0.0113,2.53)$ for $k=1.8$ , the \# symbol refers to
the point $(0.0058,3.55)$ for $k=7$ and the X symbol refers to the point (0.0054,3.9) for $k=13.2$.\label{kappa}}
\end{sidewaysfigure}

We simulated the system for three different values of  $k$ (see fig.\ref{kappa}): $k = 1.8$ ($K = W = 0.55$), $k = 7$ ($K = W = 0.85$) 
and $k = 13.2$ ($K = W = 1$).
Recall that the relation  between $\left( K,W \right)$ and $k$ is $\mbox{sinh} \left( 2K \right) \mbox{sinh} \left( 2W \right) = k$ and $k > 1$ corresponds to the broken symmetry phase.

In order to study the effects of geometry variation on nucleation
for each value of $k$ we smoothly changed the values of $K$ and $W$  from  $K = W$ to the value of $W$ and $K$ corresponding to $\xi_y = 50=\frac{L}{2}$
(i.e. reaching a quasi one dimensional geometry), we let the system thermalize in this asymmetric point and then moved it back again to the symmetric 
$K=W$ geometry. 

We checked that, for all the values of $k$ that we studied, the results were essentially independent from the velocity of this transformation provided the
thermalization time in the asymmetric points was large enough. Thus for simplicity we shall only report the results corresponding to two simulations protocols.

The first one (protocol I) corresponds to a stepwise change in the value of
the couplings. More precisely we started with $5\times 10^3$ iterations with $K=W$ in order to thermalize the system, we then changed in a 
stepwise way the couplings
to the values corresponding to the maximal asymmetry (last column of  tab.\ref{tab1}) and let the system thermalize 
for $5\times 10^4$ iterations. Finally, again in a stepwise manner, we change 
 the couplings back to the symmetric value and run the simulation for  $10^4$ iterations.

The second one (protocol II) corresponds to a more gradual change of the couplings. In this case, after the first $5\times 10^3$ iterations at $K=W$,
the system is
gradually squeezed by simulating it for $10^4$ iterations for each one of the three choices of asymmetric couplings reported in tab.\ref{tab2}. 
Then it is gradually moved back
to the symmetric case following the same procedure and is finally allowed to thermalize again for $10^4$ iterations at  $K=W$. 

For all the values of $\epsilon$ and $k$ that we studied we then compared our results with those obtained by keeping the couplings (hence the
geometry) unchanged (protocol III). 
 We used this third set of simulations to identify the minimum value of $\epsilon$ detectable by the system.
 Details on the parameters used in the simulations can be found in tab.\ref{tab1}.
All the three protocols run for exactly  
$65 \times 10^3$ iterations and can thus be compared among them without bias.

Due to the non trivial shape of the external magnetic field, besides the usual magnetization $M = \frac{1}{N}\sum_i S_i$
 we used  the following order parameter ~\cite{fcgc07} to measure the polarization degree of the system:
\begin{equation} 
\label{sigma}
\sigma= \frac{1}{2} \frac{\sum_{i}^{N} \left( c_i - c \right) S_i}{\sum_{i}^{N} \left| c_i - c \right| } 
\end{equation}
that is
\begin{equation} 
\sigma =\frac{1}{2} \frac{\sum_{i}^{N} \epsilon \mbox{cos} \left( \frac{\pi x_i}{L_x} \right) S_i}{\sum_{i}^{N} \left| \epsilon \mbox{cos} \left( \frac{\pi x_i}{L_x} \right) \right|} .
\end{equation}

\begin{sidewaysfigure}
  \centering
  \includegraphics[scale=1.3]{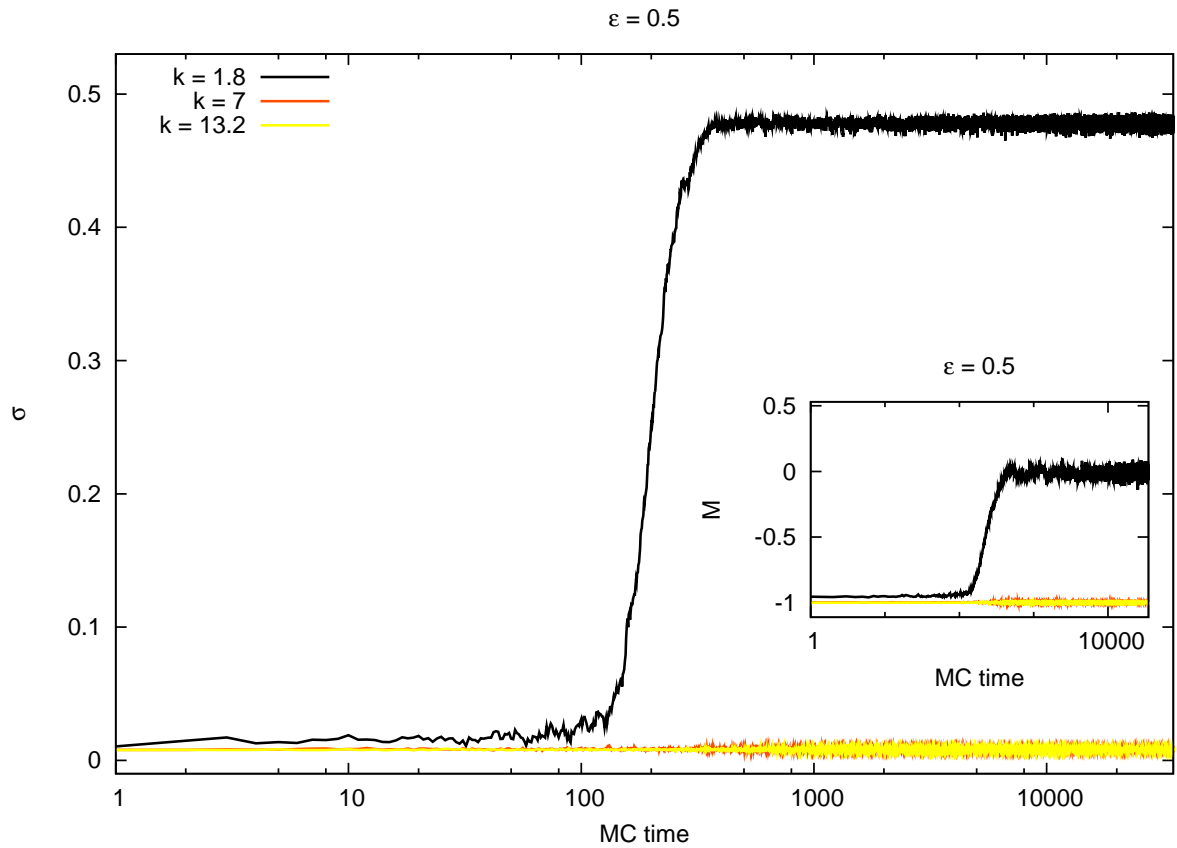}
  \caption{
  (Color online). Order parameters $\sigma$ and $M$ as functions of Monte Carlo time for three values of $k$, $\epsilon = 0.5$  and symmetrical couplings 
  $K=W$. From the top to the bottom of the figure $\sigma$ and $M$ for $k = 1.8$ in black, for $k=7$ in orange (grey) and for $k=13.2$ in yellow (light grey).\label{sigma+mag05}}
\end{sidewaysfigure}

\begin{sidewaysfigure}
  \centering
  \includegraphics[scale=0.7, angle=270]{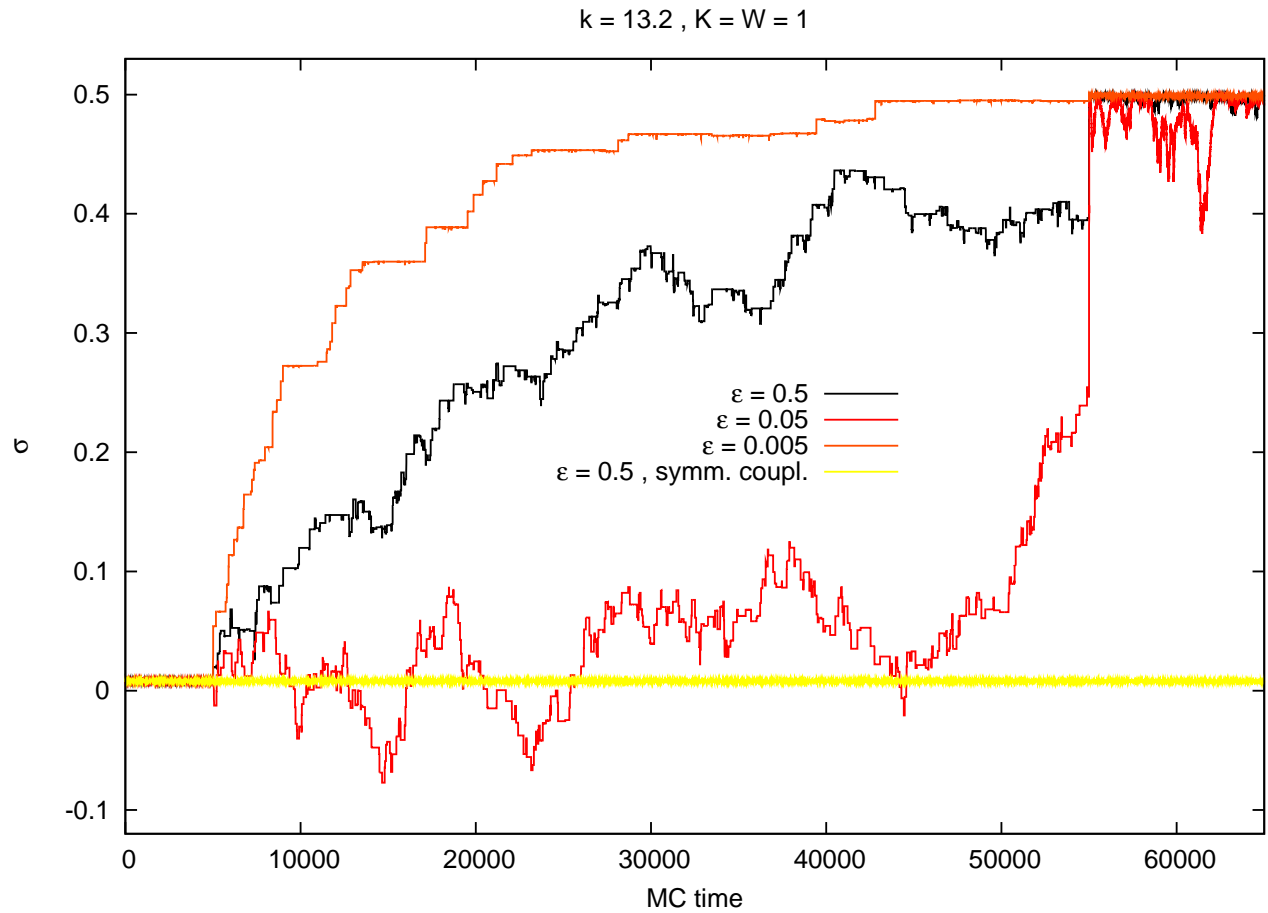}
  \caption{(Color online). Order parameter $\sigma$ as function of Monte Carlo time for $k = 13.2$ and several values of $\epsilon$ for a simulation following protocol I. From the top to the bottom of the figure, in orange (grey) data for $\epsilon = 0.005$, in black for $\epsilon = 0.5$, in red (dark grey) for $\epsilon = 0.05$ and in yellow (light grey) for $\epsilon = 0.5$ and symmetrical couplings. \label{sigma-rough}}
\end{sidewaysfigure}

\begin{sidewaysfigure}
  \centering
  \includegraphics[scale=0.7, angle=270]{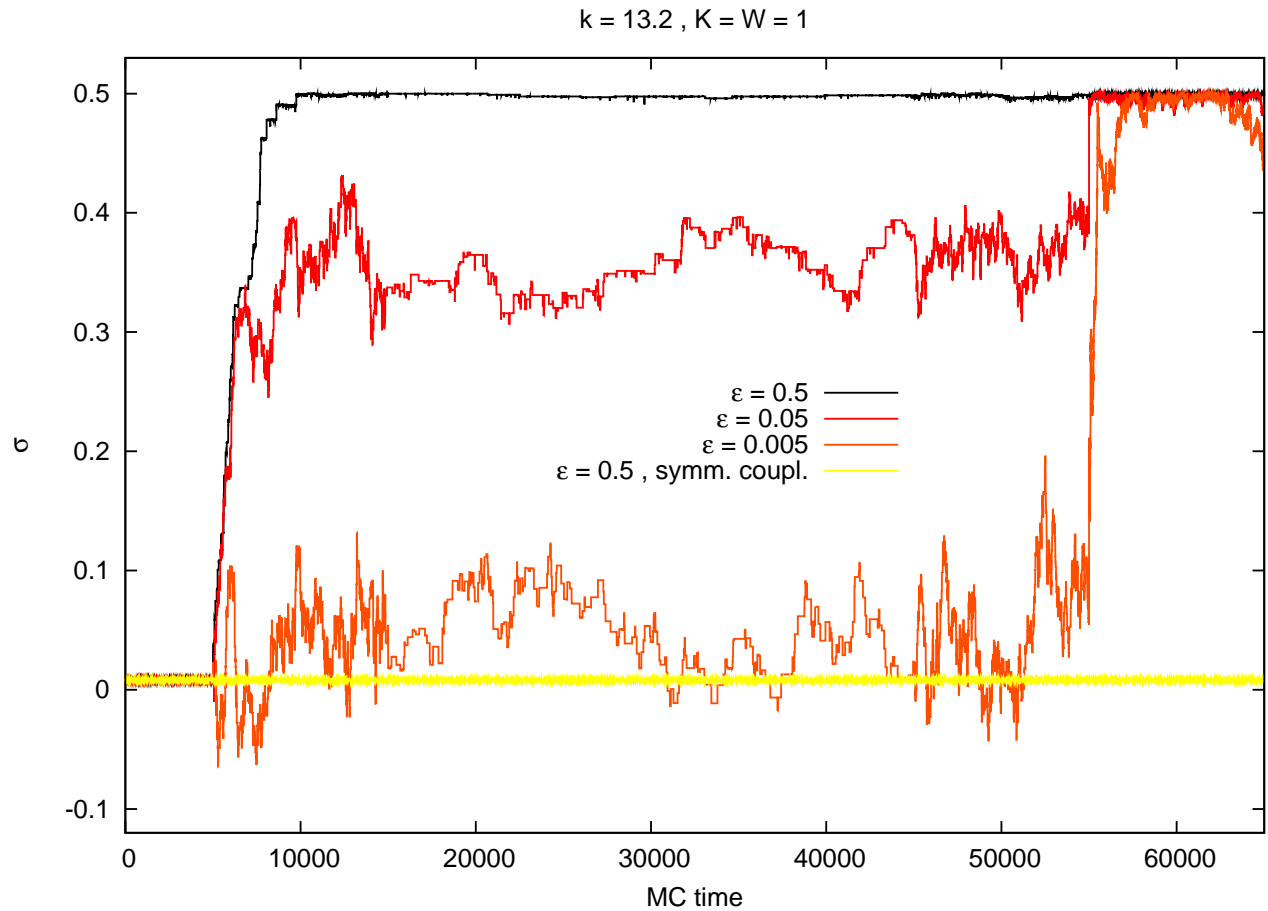}
  \caption{(Color online). Order parameter $\sigma$ as function of Monte Carlo time for $k = 13.2$ and several values of $\epsilon$ for a simulation following protocol II. From the top to the bottom of the figure, data for $\epsilon = 0.5$ in black, for $\epsilon = 0.05$ in red (dark grey), 
for $\epsilon = 0.005$ in orange (grey) and for $\epsilon = 0.5$ and symmetrical couplings in yellow (light grey). \label{sigma-smooth}}
\end{sidewaysfigure}

The results of our simulations are summarized in tab.\ref{tab2} and figs.(\ref{sigma+mag05},\ref{sigma-rough},\ref{sigma-smooth}). 
Let us discuss our main findings.
\begin{itemize}
\item
In fig.(\ref{sigma+mag05}) we report the behaviour of $\sigma$ for three values of $k$ and $\epsilon=0.5$ 
as a function of the Montecarlo time, keeping $K=W$ (protocol III).
When $K = W$, for a fixed value of $\epsilon$ the lifetime of the metastable state increases exponentially as a function of $k$. 
This is a well known result \cite{rg94} and is
in agreement with classical nucleation theory. In fact as $k$ increases the interfacial tension (and hence the free energy $F_c$ of the critical droplet)
increases. For the smallest values of $\epsilon$ and the largest values of $k$ this time is much larger than our simulation time and as a
consequence for these values of $k$ and $\epsilon$ we did not observe a decay of the metastable state  (see tab.\ref{tab2}). In our biological interpretation 
these values of $\epsilon$ are below the detectability threshold for the cell.

\item
In figs.(\ref{sigma-rough}) and (\ref{sigma-smooth}) we report the behaviour of $\sigma$ for $k = 13.2$ and several values of $\epsilon$ 
as a function of the Montecarlo time for simulations following protocols I and II respectively.
By squeezing the lattice we see that the order parameter $\sigma$ for the three values of $\epsilon$ reported
gradually moves from $ \sim 0$ to $\sim 0.5$: i.e. at the end of the simulation  the spin up cluster 
is exactly localized around the maximum of the magnetic field. We also report for comparison the result of the simulation with no change in the couplings. In this
case due to the rather high value of $k$ the metastable lifetime is exponentially long and in fact the order parameter keeps the value $\sigma=0$ for the whole
simulation time. By gradually decreasing $\epsilon$ we could identify as $\epsilon=0.0005$ 
the threshold at which even with the squeezing protocol the system could not detect the external magnetic gradient. This value should 
be compared with the analogous value $\epsilon=0.5$ in absence of squeezing and shows that we have an enhancement of
more than three orders of magnitude in the sensitivity of the system to external stimuli.
\end{itemize}

\begin{sidewaysfigure}
  \centering
  \includegraphics[scale=0.9]{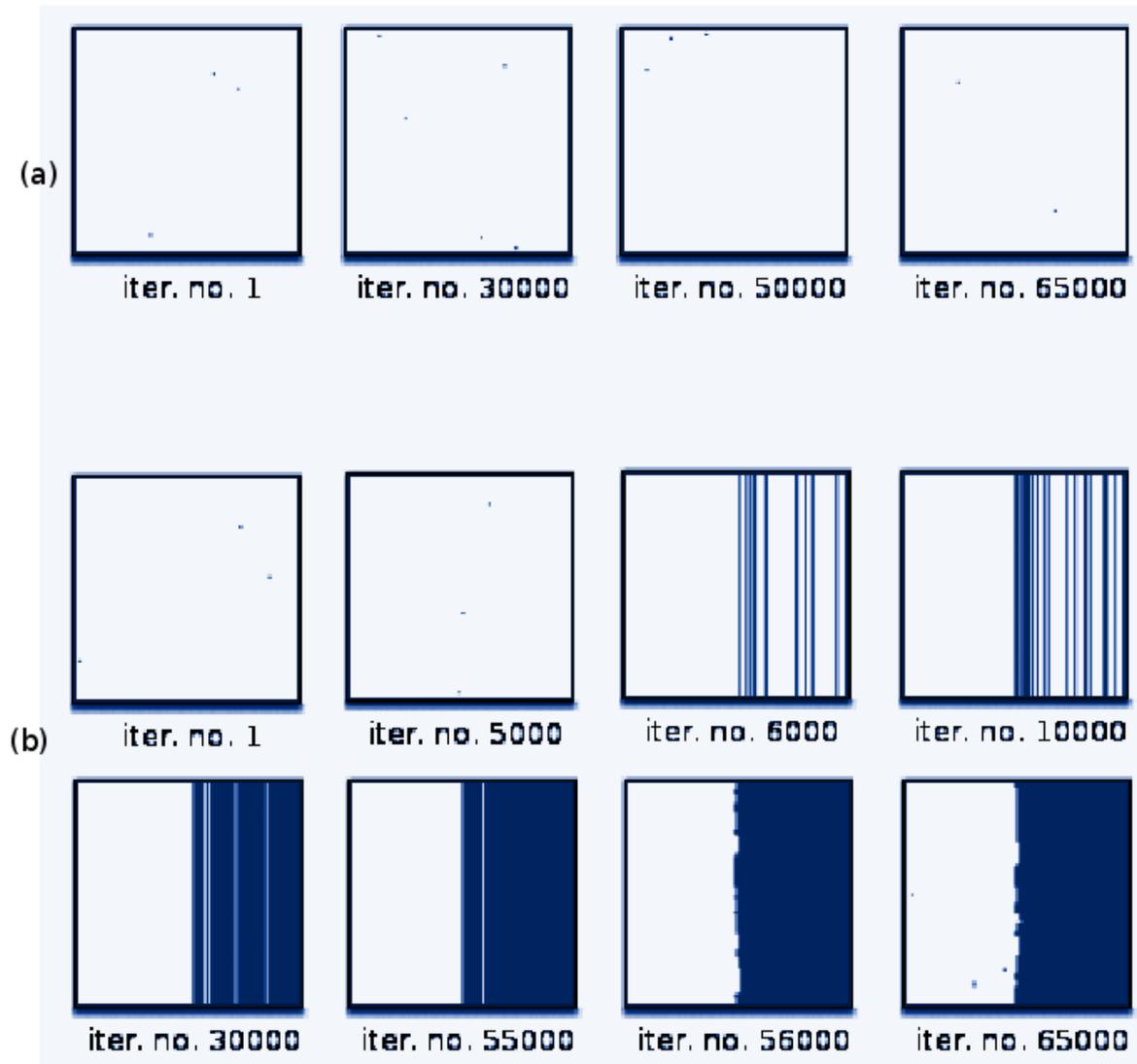}
  \caption{(Color online). The figure shows several snapshots of a square lattice with side $L=100$ following protocol III (panel (a)) and protocol I (panel (b)). Each snapshot is taken at a particular Monte Carlo step, indicated by the number of the corresponding iteration (iter.no.1, iter.no.30000 and so on) below each square lattice. Blue (black) dots are up spins.
In the panel (a), lattice with $K=W$ for the whole simulation, $k = 13.2$ and $\epsilon = 0.5$. In the panel (b), coarsening dynamics 
in presence of asymmetrical couplings between $5\times 10^3$ and $55 \times 10^3$ iterations (from the second to the fifth lattice), $k= 13.2$ and $\epsilon = 0.5$.\label{reticoli}}
\end{sidewaysfigure}

The observed behaviour can be intuitively understood as follows. By changing the couplings in protocols I and II we change the value of the interfacial tension 
in the horizontal versus the
vertical directions. This leads to the creation of asymmetric domains (vertical strips with our protocols) which play the role of nucleation drops. For large enough
asymmetry these domains can wind around the vertical direction due to the periodic boundary conditions, thus leading to an effective dimensional reduction.
In this regime the model behaves effectively as a one dimensional Ising model and the vertical strips start to aggregate as it happens for the spin kinks in the 1d
Ising model. This process is driven by the external field and can be very slow if the gradient $\epsilon$ is small. However, even if it may not suffice to completely
polarize the lattice (see the shape of the $\sigma$ parameter for the lowest $\epsilon$ values in figs.(\ref{sigma-rough}) and (\ref{sigma-smooth})), it is enough to
destabilize the original metastable configuration thus allowing the system, when it is moved back to the symmetric values of couplings, to reach the stable, polarized
state. This is well described by fig.(\ref{reticoli}) in which snapshots of the spin configurations are plotted at different time during the
simulation, both in the symmetric case and following protocol~II.

\section{Concluding remarks} 
\label{conclusions-prova} 
Recent studies \cite{g1, g2, g3} pointed out that chemotacting eukaryotic cells react to external chemical signals firstly polarizing at 
level of their inner membrane: 
the interplay between the enzymes $PTEN$ and $PI3K$ in the cytoplasm leads to the formation of two complementary clusters of phospholipids 
($PIP_2$ and $PIP_3$) on the membrane. 
If the concentration of chemoattractant is larger than a threshold value, then the $PIP_3$-rich cluster is localized around the maximum of the signal, 
determining internal cellular polarity.
The above polarization is reached thanks to a phase separation process which can be cleverly mapped on a 2d-Ising-like model mimicking a spherical 
cell \cite{fcgc07}.
However, dynamical geometry alterations play a pivotal role during the phase of signal caption. In particular it is well known that several types of eukaryotic cells
use {\itshape filopodia}, which are quasi-one-dimensional finger-like 
actin protrusions exposed at the leading edge of migrating cells,  as sensors of local external environment. 
As very subtle directional antennas, they are sofisticated tools which allow the cell to sense teeny quantity of signal 
otherwise negligible because below the threshold of detectability for a rigid spherical cell. 

In this paper, in order to describe the phase separation process 
occuring on the membrane of these filopodia we slightly modified the model proposed in \cite{fcgc07} so as to be able to define it 
on a cylindrical lattice with a radius which may change 
 as a function of time. In order to keep the exact solvability of the model we concentrated in the $N\to\infty$ limit (see above) which corresponds to the
 standard anisotropic Ising model. 
 
The main assumption behind our approach is that the phase separation scenario proposed in \cite{fcgc07}  for
the spherical cells holds 
also for filopodia, despite their extreme geometry. This assumption is supported by a few recent works on neuronal polarization showing that the $PI3$-kinase
and its phospholipid product $PIP_3$ are involved in the dendritic filopodial motility \cite{menager, luikart}.

In particular clear evidences of a phase separation process were observed in \cite{luikart}, where the autors found 
that focal accumulation of $PIP_3$ accompanies filopodial motility.     
In agreement with these findings we 
showed that squeezing the geometry of the cylinder (i.e. elongating the filopodia) allows to speed up the nucleation 
time even in presence of very small energy gaps between the stable and the metastable states.

Altogether our analysis and these experimental results support the idea that eukaryotic cells use 
conformational changes of their membrane, and in particular the protrusion of filopodia, as a tool to optimize signal detection
and speed up chemotaxis even in presence of very small
amounts of chemoattractants by taking advantage of the enhanced efficiency of the phase separation process in a quasi-one-dimensional geometry.

\vskip1.0cm {\bf Acknowledgements.}

The authors would like to thank A.Gamba, L.Grassi, T.Ferraro, G.Serini, M.Osella and L.Castagnini
for  useful discussions and suggestions. 

\newpage
\begin{sidewaystable}
\centering
\begin{tabular}{||p{1.7cm}||*{6}{c|}|}
\hline
			&		  &			&		  &		&		&		\\
$\;\;\;\;\;\; \epsilon$ &	5	  &	1		&	0.5	  &	0.05	&	0.005	&	0.0005	\\
			&		  &			&		  &		&		&		\\
\hline
\hline
			&		  &		        &		  &		&		&		\\
$k=$1.8			& $\star \bullet$ & $\star \bullet$	& $\star \bullet$ & $\bullet$	& $\otimes$	& $\otimes$	\\
			&		  &			&		  &		&		&		\\
\hline
			&		  &			&		  &		&		&		\\
$k=7$			& $\star \bullet$ & $\star \bullet$	& $\bullet$	  & $\bullet$	& $\otimes$	& $\otimes$	\\
			&		  &			&	          &		&		&		\\
\hline
			&		  &			&		  &		&		&		\\
$k=13.2$		& $\star \bullet$ & $\bullet$		& $\bullet$       & $\bullet$	& $\bullet$	& $\otimes$	\\
			&		  &			& 		  &		&		&		\\
\hline
\end{tabular}
\caption{\label{tab2}
Summary of the system behaviour for each value of $k$ and $\epsilon$. Legend: $\star$ phase separation achieved with symmetrical couplings;
$\bullet$ phase separation achieved with asymmetrical couplings; $\otimes$ phase separation not achieved during the simulation time, either with 
asymmetrical couplings.}
\end{sidewaystable}
\newpage
\begin{sidewaystable}[t]
\centering
\begin{tabular}{||p{1.3cm}||*{4}{c||}}
\hline
\multicolumn{5}{||c||}{$k=1.8$}	     			   		   								     \\ 			 
\hline
\hline
				    &			    &				&			   &		             \\	
 $\;\;\;  K$		            &  0.55 ($\xi_x=1.225$) &  0.0445 ($\xi_x=0.85$)    & 0.0225 ($\xi_x=0.85$)    & 0.0113 ($\xi_x=0.85$)    \\
				    &			    &				&			   &			     \\		
\hline
				    &			    &				&			   &			     \\
 $\;\;\;  W$       		    &  0.55 ($\xi_y=1.225$) &  1.85 ($\xi_y=12.5$)      & 2.19 ($\xi_y=25$)        & 2.53 ($\xi_y=50$)       \\
				    &			    &				&			   &			     \\
\hline
\multicolumn{5}{||c||}{$k=7$}	     			   										     \\ 	
\hline
				    &			    &				&			   &			     \\		
 $\;\;\;  K$       		    & 0.85 ($\xi_x=0.375$)  &  0.0234 ($\xi_x=0.25$)    & 0.0116 ($\xi_x=0.25$)    & 0.0058 ($\xi_x=0.25$)    \\
				    &			    &				&			   &			     \\
\hline
				    &			    &				&			   &			     \\
 $\;\;\;  W$			    & 0.85 ($\xi_y=0.375$)  &  2.85 ($\xi_y=12.5$)      & 3.20 ($\xi_y=25$)        & 3.55 ($\xi_y=50$)       \\
				    &			    &				&			   &			     \\
\hline
\multicolumn{5}{||c||}{$k=13.2$}	     			   		   							     \\ 	
\hline
				    &			    &				&			   &			     \\
 $\;\;\;  K$			    &  1 ($\xi_x=0.29$)     &  0.0209 ($\xi_x=0.2$)     & 0.0108 ($\xi_x=0.2$)     & 0.0054 ($\xi_x=0.2$)    \\
				    &			    &				&			   &			     \\
\hline
				    &		            &				&			   &			     \\
 $\;\;\;  W$			    &  1 ($\xi_y=0.29$)     &  3.22 ($\xi_y=12.5$)      & 3.55 ($\xi_y=25$)        & 3.9 ($\xi_y=50$)        \\
				    &			    &				&			   &			     \\	
\hline
\end{tabular}
\caption{\label{tab1}
Summary of the details on the parameters used in the simulations. For each value of $k$, the $K$ and $W$
values are reported together with the corresponding correlation lengths $\xi_x$ and $\xi_y$.}
\end{sidewaystable}
\newpage
\newpage
\newpage
\newpage
\newpage
\newpage
\newpage

\end{document}